# The Changing Perception of the Solar System

Authors: D. Nesvorny[1,2]*, F. Roig[2]

**Affiliations:**

[1]Southwest Research Institute, 1050 Walnut St., Suite 300, Boulder, CO 80302, USA

[2]National Observatory, General José Cristino 77, Rio de Janeiro, RJ 20921-400, Brazil

*Correspondence to: davidn@boulder.swri.edu

**Abstract**:

The solar system has changed dramatically since its birth, and so did our understanding of it. A considerable research effort has been invested in the past decade in an attempt to reconstruct the solar system history, including the earliest stages some 4.5 billion years ago. The results indicate how several processes, such as planetary migration and dynamical instabilities, acted to relax the orbital spacing of the outer planets, and provided the needed perturbation to explain the present planetary orbits that are not precisely circular and coplanar. Here we highlight this work and illustrate the key results in a computer simulation that unifies several recently developed theories. The emerging view represents another step away from the initial perception of the solar system as part of unchanging heavens.

**Summary:**

Scientists begin to realize how past planetary migration and dynamical instabilities gave the solar system many of its present attributes.

**Main text:**

The motion of planets has inspired human thought since antiquity. As seen from the Earth, the paths of planets on the sky are intriguing. They move forward for a while, then loop back, as if their projected motion is observed from a moving platform. The heliocentric world view of Copernicus emerged from a realization that all planets, including the Earth, orbit around the Sun. Next, the clockwork-like theory of epicycles, which struggled to account for observations of planetary orbits by adding more gears on the top of the previous ones, was replaced by a simpler and more accurate scheme with planets moving on ellipses (the famous first law of Johannes Kepler). This closed the deal, or it at least seemed that way, because the model accurately explained observations.

Three-quarters of a century later, Isaac Newton revealed a deeper truth in the solar system workings. He discovered that the elliptic motion of planets, among other things, can be explained by the attractive force of gravity. He also reasoned that the Kepler ellipse is only an approximation of the true nature of planetary motion. Indeed, the irregularities produced by the gravitational interaction among planets were observed, and ingenuous methods were later developed to compute them from the Newton laws. The field of celestial mechanics was born. The most celebrated triumph of the celestial mechanics became the theoretical prediction of an unseen planet, Neptune, from the deviations observed in the orbital motion of Uranus. That was in 1846.

In modern times, much research is directed toward understanding the solar system origins with the planetary orbits being considered as important clues. The elliptic nature of orbits is measured by a parameter known as the eccentricity (values >0 indicate elongation), and their tilt to a reference plane by the inclination. Mercury has the largest orbital eccentricity (e=0.17) and largest inclination (i=7 deg) among the solar system planets. Jupiter's orbit is more circular and coplanar (e=0.05 and i=0.4 deg). In principle, these values could have surged from some turbulent and difficult-to-characterize processes during the earliest stages of planet formation. The recent studies suggest, instead, that planets could have emerged from the dispersing circumsolar disk on precisely circular orbits in a common plane, and the e and i values observed today were established later [1-3].

The results can be conveniently illustrated with a computer simulation. Its starting point is the early solar system with the terrestrial and outer planets on the circular and coplanar orbits. The outer planets are assumed to be closer to the Sun than they are now, with Neptune at 20 AU and Saturn's orbital period being only 1.5 times longer than Jupiter's (this ratio is nearly 2.5 today). These assumptions are motivated by the orbital evolution of planets during the previous stage, when they exchanged the orbital momentum with a protoplanetary gas nebula, and converged inward and toward each other [4]. The gas nebula dissipation was also expected to damp orbital eccentricities and inclinations. A massive disk of small icy bodies (planetesimals) is placed beyond the orbit of Neptune. Its remains survived in the trans-Neptunian region to this day, and after Gerard Kuiper who imagined its existence in the early fifties, it is known as the Kuiper belt.

A number of things happens as the system evolves. Planetesimals leak from the outer disk onto Neptune-crossing orbits and are subsequently scattered inward or outward during close encounters with Neptune. The ones scattered outward come back and are scattered inward, where they encounter Uranus and Saturn. These planets act in much the same way as Neptune, eventually handling bodies to Jupiter, which ejects them from the solar system. The conservation of orbital momentum dictates, as planetesimals move from the outer disk inward, that Saturn, Uranus and Neptune must move outward. This process is known as the planetesimal driven migration or PDM [5, 6]. The PDM explains how the outer planets reached their present orbital radii with Neptune at 30 AU. This cannot be the end of story, because the planetary eccentricities and inclinations remain small during the PDM.

Interestingly, when Neptune reaches roughly 28 AU in the specific simulation discussed here [7], a dynamical instability develops with the inner ice giant evolving onto an orbit intersecting those of Jupiter and Saturn. The instability trigger is related to the gravitational resonances encountered by the migrating planets [1, 8]. The subsequent planetary encounters have several consequences. First, they excite eccentricities and inclinations of the outer planets to values comparable to the present ones. Second, the semi-major axes of planets evolve discontinuously during encounters, with Jupiter's semi-major axis changing by as much as 0.5 AU (so-called jumping Jupiter [9, 10]). Third, the inner ice giant is ejected into interstellar space. It is not known how many ice giants formed in the solar system, but the instability calculation with one extra planet on an initial orbit between Saturn and Uranus gives the best results (Figure 1, [7]).

As the outer solar system reconfigures, the inner planets follow the suit. If Jupiter slowly migrated due to the PDM, gravitational resonances between the terrestrial planets and Jupiter would have plenty of time to act. They would disrupt the terrestrial system orbits, eventually leading to planet-planet collisions [11]. Jumping Jupiter solves this problem, because the resonant effects are reduced when Jupiter's orbit changes discontinuously. Nevertheless, the eccentricities and inclinations of the terrestrial planets become excited (Figure 1). Most notably, in the successful simulation highlighted here, Mercury's eccentricity and inclination reach their present values. Not everything is perfect, however. For example, the orbital inclination of Mars ends up slightly lower than its present value (4 deg). This may imply that Mars had some orbital inclination initially, or that the specific evolution discussed here is still missing some important component.

Much of the future research will be directed toward the goal of improving the results shown in Figure 1. A fundamental difficulty with these efforts is that the orbital evolution during the instability is chaotic and must therefore be studied statistically. The small bodies, such as the asteroids and Kuiper belt objects, place important constraints on the evolution history of planets. While the jumping-Jupiter model with an extra ice giant owns much of its success to matching the basic properties of these reservoirs, getting things right in detail may be difficult. While we reflect on these new challenges, we cannot help from seeing the solar system in a very different light than our distant ancestors. We perceive it as an evolved physical system that reached its present state after several notable events took place during its tremulous past.

**Acknowledgments:**

D.N. acknowledges support from the NASA Origins of Solar Systems program.

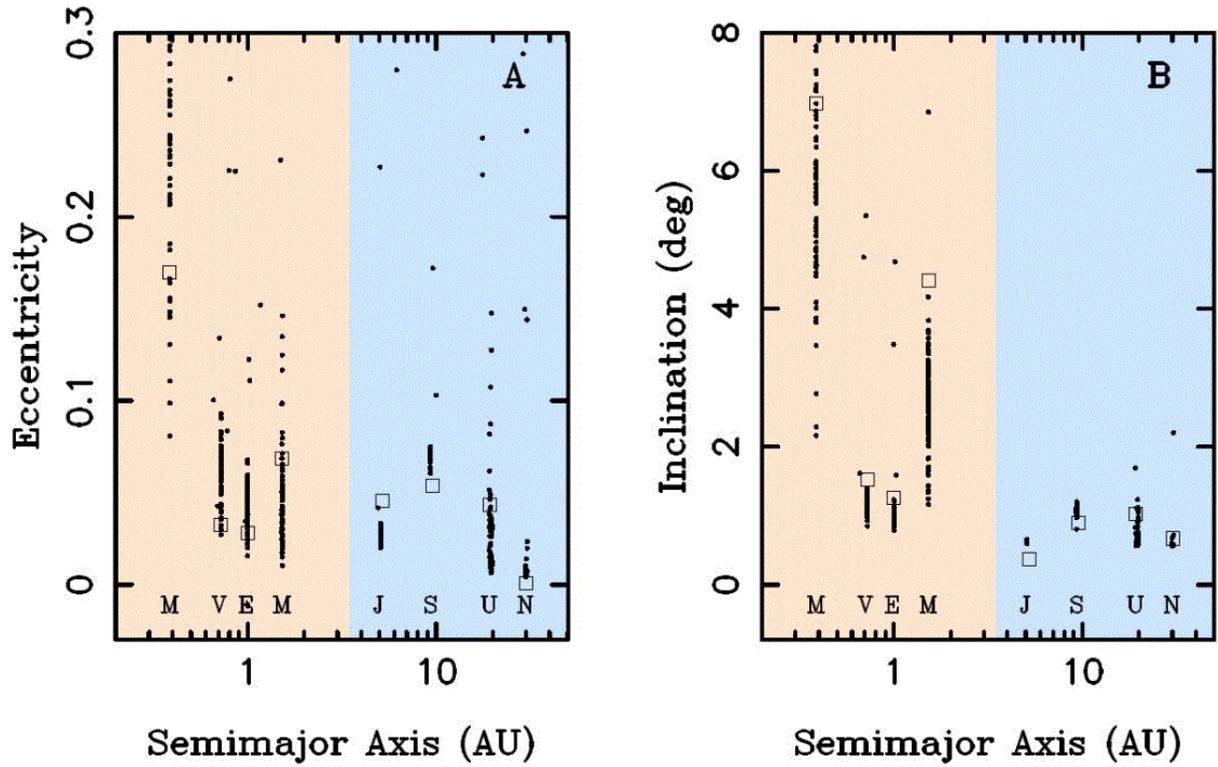

**Fig. 1**. A computer model matches the present orbits of the solar system planets. The simulation starts with the circular and coplanar orbits of all planets. An outer disk of small planetesimals with the total mass of 20 Earth masses is placed beyond the initial orbit of Neptune. The disk is the cause of planetary migration and instability, and the source of the Kuiper belt objects. The final planetary orbits (dots) obtained in a hundred of computer simulations (starting from slightly different initial conditions) match the general properties of the present orbital architecture (squares). A few small differences can be noted.